\documentclass[12pt]{article}
\usepackage{epsfig}
\usepackage{amsfonts}
\usepackage{dsfont}
\usepackage{times}

\newcommand{\nat}{{\mathds N}} 
\newcommand{\real}{{\mathds R}} 

\def\+{{+\!\!\!+}}

\newcommand{\C}{{\mathds C}}

\def\pmb#1{\setbox0=\hbox{#1}%
\kern.0em\copy0\kern-\wd0 
\kern-.04em\copy0\kern-\wd0 
\kern.08em\copy0\kern-\wd0 
\kern-.04em\raise.0433em\box0 }         
 
\newcommand{\nc}{\newcommand} 
\nc{\beq}{\begin{equation}} 
\nc{\eeq}[1]{\label{#1}\end{equation}} 
\nc{\ber}{\begin{eqnarray}} 
\nc{\eer}[1]{\label{#1}\end{eqnarray}} 
\nc{\pek}[1]{\cite{#1}} 
\nc{\enr}[1]{(\ref{#1})} 
\nc{\kal}[1]{{\cal{#1}}} 
\nc{\dott}{\;\cdot\;} 

\def\0 {\nonumber}

\begin{document} 
\setcounter{page}{0}
\newcommand{\inv}[1]{{#1}^{-1}} 
\renewcommand{\theequation}{\thesection.\arabic{equation}} 
\newcommand{\be}{\begin{equation}} 
\newcommand{\ee}{\end{equation}} 
\newcommand{\bea}{\begin{eqnarray}} 
\newcommand{\eea}{\end{eqnarray}} 
\newcommand{\re}[1]{(\ref{#1})} 
\newcommand{\qv}{\quad ,} 
\newcommand{\qp}{\quad .} 

\def\qp{Q_+}
\def\qm{Q_-}
\def\qbp{\bar Q_+}
\def\qbm{\bar Q_-}
\def\sgh{\Sigma_{g,h}}

\begin{titlepage} 
\begin{center} 

\hfill SISSA 55/2007/EP\\  
                         
\vskip .3in \noindent 


{\Large \bf{The holomorphic anomaly for open string moduli}} \\

\vskip .2in 

{\bf Giulio Bonelli and Alessandro Tanzini}

\vskip .05in 
{\em\small International School of Advanced Studies (SISSA) and INFN, Sezione di Trieste\\
 via Beirut 2-4, 34014 Trieste, Italy} 
\vskip .5in
\end{center} 
\begin{center} {\bf ABSTRACT }  
\end{center} 
\begin{quotation}\noindent  
We complete the holomorphic anomaly equations for topological strings 
with their dependence on open moduli. 
We obtain the complete system by standard path integral arguments generalizing 
the analysis of BCOV (Commun.\ Math.\ Phys.\  {\bf 165} (1994) 311)
to strings with boundaries.
We study both the anti-holomorphic dependence on open moduli
and on closed moduli in presence of Wilson lines.
By providing the
compactification \`a la Deligne-Mumford of the moduli space of Riemann
surfaces with boundaries, we show that the open holomorphic anomaly 
equations are structured on the (real codimension one) boundary components
of this space.
\end{quotation} 
\vfill 
\eject

\end{titlepage}

\tableofcontents

\section{Introduction}

The holomorphic anomaly equations \cite{bcov} are a most powerful tool which
potentially allows for the complete solution of topological string theories 
\cite{Witten:1991zz}, once complemented with suitable methods to fix the holomorphic
ambiguities. 
Nowadays they are experiencing a second youth due to 
the development of new techniques based on modular invariance
which are very effective to solve the recursion relations and
fix the holomorphic ambiguity up to very high orders \cite{klemm}.  
Moreover, it has been possible 
to define, via string dualities,
a clear correspondence with matrix models \cite{marino}.

The most exciting and mysterious string duality in the game is the one among open and 
closed strings.
This predicts that open and closed string theories in generically different 
target space backgrounds can be mapped one into the other 
via a suitable dictionary.
Open/closed duality has to manifest in its full glory in the cases when 
complete control of the string theory is at hand. This is indeed the case 
of the topological string.
In this case, on the closed string side, the full solution of the theory should
be provided by the 
holomorphic anomaly equations (from now on HAE's for short) 
and therefore its open string dual is expected to be fully tractable too.
The considerable amount of results on topological aspects of gauge/string dualities 
obtained during the last years, starting from \cite{GV,DV}, encourage to consider
the problem of formulating HAE's for open string moduli. 
Actually, the HAE's for closed moduli in presence of boundaries 
has been recently explored in \cite{large,marino} for local CY's by exploiting
the relation with matrix models and in 
\cite{jw} for compact CY's extending the original BCOV formulation
\footnote{After the submission of this paper, the interesting twin papers
\cite{Alim:2007qj} and \cite{KM} appeared explicating and solving the extended HAE's 
of \cite{jw} for closed moduli on the quintic.}.
The boundary effects calculated in \cite{jw} have been immediately 
reinterpreted in \cite{zio}
in terms of a shift of variables in the usual BCOV equations. 
This was done at frozen open string moduli.
Moreover, various aspects of open topological string disk amplitudes 
were studied in \cite{ov,vafa-o,open} for local CY's and in \cite{panda} for compact ones. 
Some of these amplitudes have been observed to be related to four-dimensional
effective terms which are of relevance in phenomenological applications
of open superstring compactifications, as computing Yukawa couplings
\cite{Marchesano,Rodolfo} and gaugino masses \cite{ant}. Moreover
the explicit calculations of these papers display an anti-holomorphic dependence.

The aim of this letter is to start exploring the HAE's for open strings 
and the intertwining among open and closed moduli.
Our main results are two. First of all,
we formulate the HAE's for open string moduli. Their structure is modeled, 
analogously to the closed string case, on the boundary of a suitable compactification of 
the moduli space of open Riemann surfaces.
The definition of this compactification scheme at all genera is at our knowledge new.
Secondly, we complete the HAE's for closed moduli in the case in which open strings moduli are 
turned on.
We will work out our results for simplicity in the B-model language, 
but its analogue holds for the A-model too.

The plan of the paper is the following. 
In section 2 we recall some notations and list 
the marginal bulk and boundary deformations of the open B-model.
In section 3 we formulate the relevant compactification of the moduli space of open 
Riemann surfaces by generalizing the 
recipe by Deligne and Mumford \cite{DM}.
In section 4 we obtain via detailed path integral arguments the HAE's for the open 
string moduli corresponding to the marginal 
boundary deformations 
and
in Section 5 we complete the HAE's for the closed string moduli in presence of open string ones.
We left Section 6 for some comments and open questions.

\section{Boundary marginal deformations}

Let us start by defining the B-model action and path integral in the case of strings with boundary.
In the standard BCOV notation\footnote{We follow the conventions of 
\cite{mirror,klemmrev}, to which we refer for details. 
}
the action is
\be
S_{B(bulk)}=\left\{Q, V\right\} +W
\label{Bact}\ee
where $Q=\qbp+\qbm$ is the BRST charge, 
$V=\int_{\sgh} g_{I\bar J}\rho^I\wedge *dX^{\bar J}$  is the gauge fermion
and 
$W=-\int_{\sgh}\theta\cdot \wedge D\rho +\frac{1}{2} 
R\cdot (\rho\wedge\rho\,\, \eta\theta)$ is the classical action \cite{Witten:1991zz}. 
The B-model partition function at given genus $g$ and holes $h$ 
is calculated by the path integral
\be
F_{g,h}
=
\int_{\bar{\cal M}_{g,h}} 
\langle\prod_{k=1}^{3g-3+h}|(\mu_k,G^-)|^2 \prod_{a=1}^h (\lambda_a,G^-)\rangle_{\Sigma_{g,h}}
\label{Bamp}\ee
where
$\bar{\cal M}_{g,h}$ is the (compactified) moduli space of complex structures over 
Riemann surfaces $\Sigma_{g,h}$. This will be described in detail in the next section.
In (\ref{Bamp}), $\mu_k$ are the Beltrami differentials parametrizing 
the variations of the metric in the bulk of the 
Riemann surface and the positions of the boundary components, $\lambda_a$ 
are the Beltrami differentials 
associated with the variations of the lengths of the boundary components: 
as such they are supported near the boundary $\partial\sgh$ itself.
Moreover $(\mu,G^-)=\int_{\sgh}\mu^{\bar z}_z G^-_{\bar z \bar z}$
is the pairing among the $G^-$ supercurrent and the complex Beltrami differential $\mu$,
$(\lambda,G^-)= \int_{\sgh}\lambda^{\bar z}_z G^-_{\bar z \bar z}+\bar\lambda_{\bar z}^z \bar G^-_{zz}$
is the pairing among the supercurrents $G^-$ and $\bar G^-$ with the Beltrami differentials $\lambda$
corresponding to the real moduli. Finally, $\langle\dots\rangle_{\Sigma_{g,h}}$
indicates the path integral amplitude of the topological $\sigma$-model.
The structure of the supercurrent insertions paired with the relevant Beltrami differentials 
generates the Weyl-Petersson measure on $\bar{\cal M}_{g,h}$.

In the case of open strings, it is possible to add to the bulk action (\ref{Bact}) the boundary coupling 
to a gauge field in the form of a supersymmetric Wilson line. 
This reads\footnote{If the gauge bundle is non-trivial, a more refined expression is required
see \cite{rudisciu}.}
\be
S_{B(boundary)} = i\oint_{\partial\sgh} \left(X^*(A)+
\left(F_A\right)_{I\bar J}\rho^I\eta^{\bar J}\right)
\label{wl}\ee
and can be rewritten \cite{mirror} in the manifestly supersymmetric form
\be 
S_b= Q\oint_{\partial\sgh} A_I(X)\rho^I + \oint_{\partial\sgh} \bar Q A_{\bar I}(X)\eta^{\bar I}
\label{wl'}\ee
if the gauge connection is holomorphic, that is if it satisfies $F_A^{(2,0)}=0$.
In (\ref{wl'}) we used the anti-BRST charge $\bar Q=\qp+\qm$. 
The total action of the B model is therefore
\be
S_B=S_{B(bulk)}+S_{B(boundary)} \ \ .
\label{Btot}\ee
The generalization to the case of non abelian gauge bundles is straightforward and corresponds
to the usual path-ordering of the Wilson line (\ref{wl}).

The generic marginal deformations are given by the closed string moduli corresponding to variations of the 
CY complex structure and by the open string moduli corresponding to the variations of 
the complexified gauge connection.
Specifically, we have\footnote{Not to overweight the notation, we omit the summation over the boundary components 
which is left understood.}
\bea
\delta S_B =  \qbp\qbm\int_{\sgh} \delta t^{\bar i}\phi_{\bar i} +\int_{\sgh} \qp\qm 
\delta t^i\phi_i+ \nonumber\\ 
+ Q\oint_{\partial\sgh}\left(
\delta t^{\bar\alpha}\Theta_{\bar\alpha} +\delta t^{\bar i}\Psi_{\bar i}
\right)
+
\oint_{\partial\sgh}\bar Q\left(
\delta t^\alpha \Theta_\alpha + \delta t^i\Psi_i\right)
\label{mdef}
\eea
where, for the B model 
\newpage
\bea
\phi_{\bar i}&=&\left(w_{\bar i}\right)_{IJ}(X)\rho_z^I\rho_{\bar z}^J,
\quad\quad\quad \quad \quad  \ \ \ \ \ \
\phi_i=\left(\bar w_i\right)_{\bar I\bar J}(X)\eta^{\bar I}\theta^{\bar J},\\
\Theta_{\bar\alpha}&=&\left(\delta A^{(1,0)}_{\bar\alpha}\right)_I(X)\left(\rho^I_z+\rho^I_{\bar z}\right),
\quad \ \ \ \ 
\Theta_\alpha=\left(\delta A^{(0,1)}_{\alpha}\right)_{\bar I}(X)\eta^{\bar I},\\
\Psi_{\bar i}&=&\left[\left(w_{\bar i}\right)_I^{\bar J}A^{(0,1)}_{\bar J}\right](X)
\left(\rho^I_z+\rho^I_{\bar z}\right), \quad
\Psi_i=\left[\left(w_{i}\right)_{\bar I}^{J}A^{(1,0)}_{J}\right](X)\eta^{\bar I}.
\label{notation}\eea
Notice that here and in the following we use latin low-case letters for closed
string moduli $t^i$ and greek low-case letters for open string moduli $t^\alpha$.
In (\ref{notation}),
$w_{\bar i}$ is a basis of Beltrami differentials on the target space, so that 
$\delta t^{\bar i}w_{\bar i}$
parametrizes the variation of the target space complex structure,
and similarly $\delta A_\alpha^{(0,1)}$ and $\delta A_{\bar \alpha}^{(1,0)}$ 
for the variation of the complexified gauge connection. Notice that, as it is clear from (\ref{mdef}),
the complex moduli couple to the boundary action but the contrary doesn't hold.
This simple fact has profound consequences on the structure of the complete holomorphic anomaly equations.

In order to complete the holomorphic anomaly equations with variations of the open strings moduli, we have to 
study the compactification of the moduli space of Riemann surfaces with boundaries.

\section{Compactification of the moduli space of Riemann surfaces with boundaries}

Let $\sgh$ be as above and let us consider the moduli space 
${\cal M}_{g,h}$ of 
inequivalent complex structures over it.
We take $\sgh$ to be equipped with a constant curvature metric 
with vanishing geodesic curvature along the boundary components.
In this section we formulate the analog of the Deligne-Mumford 
compactification of ${\cal M}_{g}$ for the case at hand.
In the boundaryless case, the set of Riemann surfaces is augmented by 
the inclusion of surfaces with nodes in order to stabilize 
the shrinking to zero length of closed 1-cycles.
If boundaries are present, the situation can be treated similarly
by adding boundary nodes.
In fact, these are generated by shrinking to zero length open $1$-cycles 
with end points on the boundary.
This means that we have to consider the full set of Riemann surfaces
with marked points in $\sgh\setminus\partial\sgh$, which are the usual 
ones, as well as marked points on the boundary $\partial\sgh$.

Let us denote by 
${\cal M}_{g,h,n,{\tt m}}$ 
the moduli space of Riemann surfaces with
genus $g$,
$h$ holes, $n$ marked points in $\sgh\setminus\partial\sgh$
and ${\tt m}\in \nat^h$ ordered marked points on the $h$ boundary components.
If the Euler characteristic\footnote{This formula can be obtained straightforwardly
just by building the Schottky double of the Riemann surface with nodes and then assigning 
democratically among the two halves the weight of the boundary punctures.}
\be
\chi=2-2g-n-h-\frac{1}{2}|{\tt m}|
\label{euler}\ee
is negative, then the real dimension of such a space is
\be
dim_{\real}{\cal M}_{g,h,n,{\tt m}}=6g-6+2n + 3h + |{\tt m}|,
\label{dimension}\ee 
where $|{\tt m}|=\sum_{a=1}^h m_a$
is the total number of boundary punctures.

The boundary components of ${\cal M}_{g,h,n,{\tt m}}$ can be reached 
by two distinct limiting procedures, that is by 
shrinking to zero length 
homotopically non trivial 
closed paths or open paths ending on the boundary.
These procedures generate different boundary components which are generically of different codimensions.
The Euler characteristic \ref{euler} is stable under these degenerations.
Let us describe them in detail.
\begin{figure}[!ht]
\begin{center}
\includegraphics{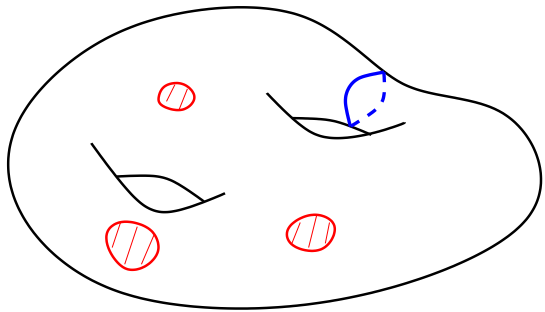}
\caption{boundary 
associated to a non-dividing closed path}
\label{pinching-closed}
\end{center}
\end{figure}
\begin{figure}[!ht]
\begin{center}
\includegraphics{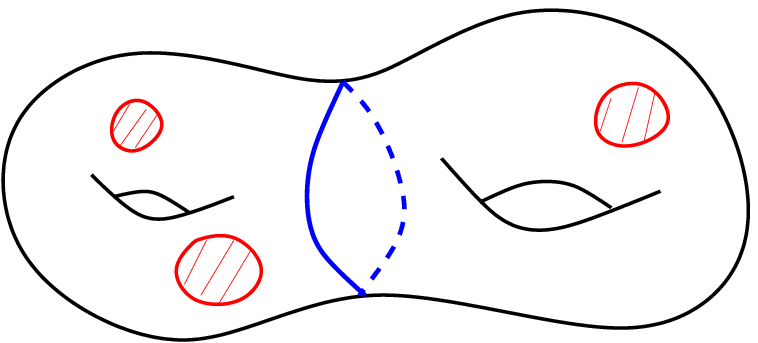}
\caption{boundary 
associated to a dividing closed path}
\label{dividing-closed}
\end{center}
\end{figure}
\begin{figure}[!t]
\begin{center}
\includegraphics{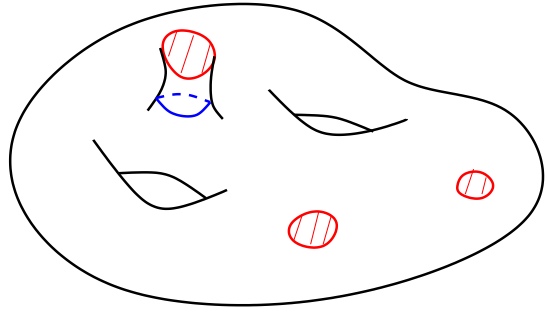}
\caption{boundary 
associated to the shrinking
of a hole}
\label{shrinking}
\end{center}
\end{figure}
Let us start from the case of closed paths.
In this case we have
\begin{equation}
\partial_c {\cal M}_{g,h,n,{\tt m}}={\cal M}_{g-1,h,n+2,{\tt m}}\cup 
\coprod_{
{\tiny
\matrix{
g_1+g_2&=&g\cr 
h_1+ h_2&=&h\cr 
n_1+n_2&=&n+2\cr 
{\tt m}_1\oplus{\tt m}_2&=&{\tt m}}
}}
{\cal M}_{g_1,h_1,n_1,{\tt m}_1}\times {\cal M}_{g_2,h_2,n_2,{\tt m}_2} 
\label{cb}\end{equation}
where the first component corresponds to a non dividing cycle, see Fig. \ref{pinching-closed},
and the others to dividing ones, see Fig. \ref{dividing-closed}.
In the above sum also genus zero contributions are counted.
In particular, if the closed path encircles a single hole as in Fig. \ref{shrinking}, 
then the resulting boundary component is the zero length limit of the 
hole and its real codimension is equal to one.
\begin{figure}[!ht]
\begin{center}
\includegraphics{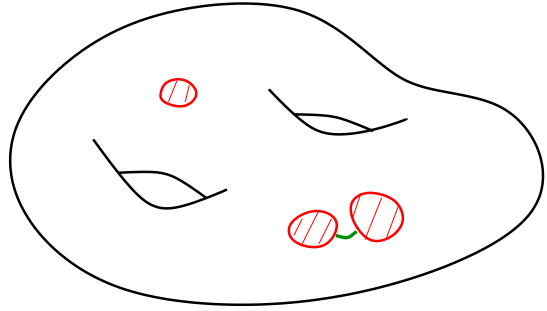}
\caption{boundary associated to colliding holes}
\label{colliding}
\end{center}
\end{figure}
\begin{figure}[!ht]
\begin{center}
\includegraphics{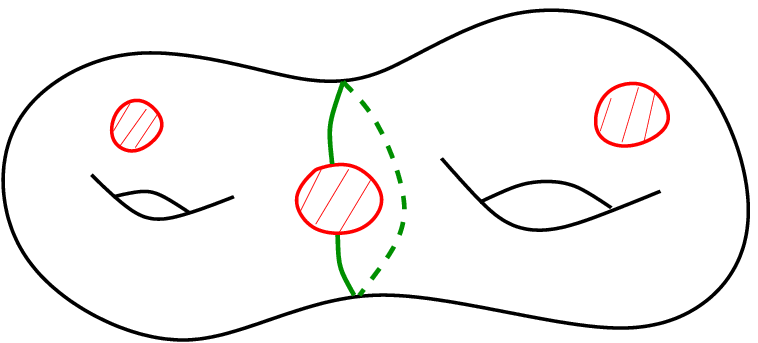}
\caption{boundary associated to a dividing open path}
\end{center}
\label{dividing}
\end{figure}
\begin{figure}[!ht]
\begin{center}
\includegraphics{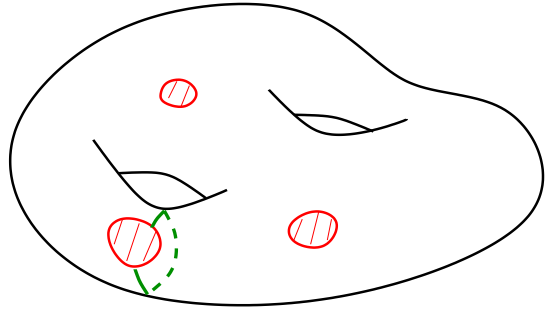}
\caption{boundary associated to a non-dividing open path}
\label{pinching}
\end{center}
\end{figure}
For open paths one has two choices regarding if the path connects two distinguished or the 
same boundary components. In the first case, see Fig. 4,
the path can not be dividing, while it can be dividing 
or not dividing in the latter as shown in 
Figs. 5 and 6
respectively.


 The boundary components are then three
\bea
\partial_o {\cal M}_{g,h,n,{\tt m}}&=&
{\cal M}_{g-1,h+1,n,\hat{\tt m}\oplus(m_l+1,m_r+1)}
\cup \label{open} \\
&&\coprod_{
{\tiny
\matrix{
g_1+g_2&=&g\cr
n_1+n_2&=&n\cr
h_1+h_2&=&h+1\cr
{\tt m}_1\oplus{\tt m}_2&=&\hat {\tt m}\oplus(m_l+1,m_r+1)}
}}
{\cal M}_{g_1,h_1,n_1,{\tt m}_1}\times {\cal M}_{g_2,h_2,n_2,{\tt m}_2}
\cup
{\cal M}_{g,h-1,n,\hat{\hat{\tt m}}\oplus(m+m'+2)}
\nonumber
\eea
where the first one is for non dividing open paths connecting the same boundary component
(Fig. 6),
the second one is for dividing open paths connecting the same boundary component (Fig. 5),
and the third one is for open paths connecting different boundary components (Fig. 4).
In the above formulas, the "hats" over the boundary punctures labels
means the omission of the entry on the vector corresponding to the boundary component(s)
over which the open path ends.
Notice the important fact that all the boundary components in $\partial_o {\cal M}_{g,h,n,{\tt m}}$
are of real codimension one.

\begin{figure}[!ht]
\begin{center}
\includegraphics{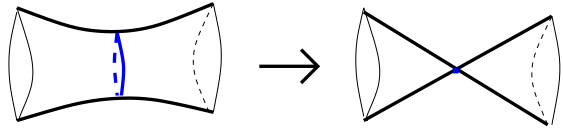}
\caption{shrinking of a closed path as $\epsilon\to0 $.}
\label{nodal}
\end{center}
\end{figure}

\begin{figure}[!ht]
\begin{center}
\includegraphics{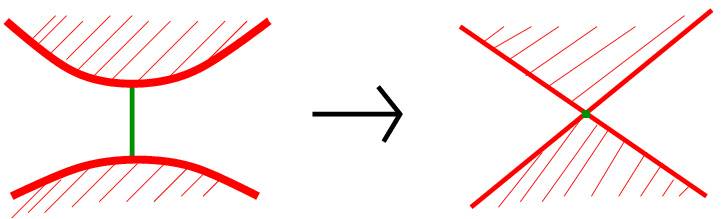}
\caption{shrinking of an open path as $\varepsilon\to 0$.}
\label{eight}
\end{center}
\end{figure}

Let us here insist on the relevance of the specific codimensionality.
Actually, in the vicinity of a closed shrinking path, the local geometry of the Riemann surface
is that of the collar $zw=\epsilon$, $\epsilon\in\C$ as $\epsilon\sim 0$. This geometry has 
a trivial $S^1$ symmetry corresponding to the twisting angle along the path of the phase of the
complex valued plumbing fixture $\epsilon$. This implies that the 
corresponding boundary component in the moduli space has 
{\it complex} codimension one\footnote{The only exception is given by the already discussed 
hole's shrinking where the $S^1$ coordinate stays as an automorphism of the punctured disk.}.
In the limiting case, one obtains the nodal geometry $zw=0$, see Fig. \ref{nodal}.

On the contrary, in the vicinity of an open shrinking path, the local geometry of the Riemann surface is
the exterior of an hyperbola $Re(z) Im(z)>\varepsilon$, $\varepsilon\in\real^+$ in the limit $\varepsilon\sim 0$.
This geometry has no $S^1$ symmetry at all and therefore the corresponding 
boundary component in the moduli space has real codimension one.
In the limiting case, one obtains locally the biquadrant geometry $Re(z) Im(z)>0$ 
corresponding to the boundary nodes, see Fig.\ref{eight}.

We will show in the following that the holomorphic anomaly for open string moduli is structured over 
the decomposition of $\partial_o {\cal M}_{g,h,n,{\tt m}}$
in the very same way as the (extended) holomorphic anomaly for closed string moduli is 
structured over the decomposition of $\partial_c {\cal M}_{g,h,n,{\tt m}}$.

\section{The open moduli holomorphic anomaly}

In this section we obtain the holomorphic anomaly equations for open moduli.
This will be done by generalizing the path integral approach of BCOV to the 
variation of the $Q$-exact part of the boundary action and by pulling the corresponding 
{\it conserved} supercharge.

Let us start by considering the boundary marginal deformations associated to the operators 
$\Theta_{\bar\alpha}$ in (\ref{mdef}). 
So we calculate
\be
\partial_{\bar t_{\bar \alpha}}F_{g,h}=
\int_{{\cal M}_{g,h}} 
\langle Q\int_{\partial\sgh}\Theta_{\bar\alpha}
\prod_{k=1}^{3g-3+h}|(\mu_k,G^-)|^2 \prod_{a=1}^h (\lambda_a,G^-)\rangle_{\Sigma_{g,h}}
\label{uno}\ee
where action of the supercharge on the boundary integral is given by
\be
Q\oint_{\partial\sgh}\Theta_{\bar\alpha}=
\oint_{\partial\sgh} dt \int_{\gamma_t} dt' \left( G^+ + \bar G^+\right)(t') 
\Theta_{\bar\alpha}(t) \ \ ,
\label{qaction}
\ee
with $\gamma_t$ the path encircling the $\Theta_{\bar\alpha}$ evaluation point as in the
following Fig.9.
\begin{figure}[!h]
\begin{center}
\includegraphics{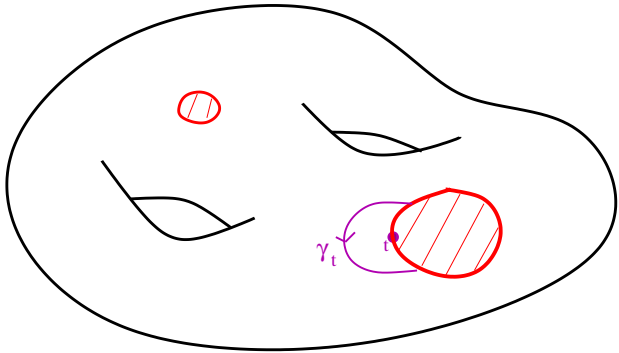}
\caption{action of the supercharge at the boundary.}
\label{encircle}
\end{center}
\end{figure}
We calculate (\ref{uno}) by pulling the supercharge $Q$ against the measure.
The supercharge $Q$ acts both on the complex and the real Beltrami differentials $(\mu,G^-)$
and $(\lambda, G^-)$ respectively. 
By using the standard superalgebra and the formula for the derivative with respect to the moduli 
$\partial_{n}\langle X \rangle=\langle X\int T\cdot\nu_n\rangle$, where $\nu_n$ is a Beltrami differential
corresponding to the generic modulus $n$, we obtain
\bea
\partial_{\bar t_{\bar \alpha}}F_{g,h}=
\int_{{\cal M}_{g,h}} 
\left\{
\sum_{j=1}^{3g-3+h}\frac{\partial}{\partial m_j}
\langle \int_{\partial\sgh}\Theta_{\bar\alpha}
(\bar\mu_j,\bar G^-)
\prod_{k\not=j}|(\mu_k,G^-)|^2 \prod_{a=1}^h (\lambda_a,G^-)\rangle_{\Sigma_{g,h}}
\right.\0 \\\left.
+\,\,\,{\rm cplx. conj.}\,\,\,
+\sum_{b=1}^h \frac{\partial}{\partial l_b}
\langle \int_{\partial\sgh}\Theta_{\bar\alpha}
\prod_{k=1}^{3g-3+h}|(\mu_k,G^-)|^2 \prod_{a\not= b} (\lambda_a,G^-)
\rangle_{\Sigma_{g,h}}
\right\} \ \ .
\label{nome}
\eea
Notice that the resulting amplitude is different to the one that is produced 
by deforming via {\it bulk} marginal 
operators $\phi_{\bar i}$ (see \cite{bcov}). In that case one has to pull
{\it two} supercharges against the measure and therefore gets {\it two}
derivatives w.r.t. moduli $\partial_m\partial_{\bar m}\langle\ldots\rangle$
picking up the logarithmically divergent term in the correlation function
$\langle\ldots\rangle_{\Sigma_\epsilon}\sim 
\langle\ldots\rangle_{\Sigma_{nodal}} ln|\epsilon| + {\rm regular~terms}$.
By varying instead via {\it boundary} marginal operators $\Theta_{\bar\alpha}$
one is pulling {\it one} supercharge against the measure and therefore gets
a single derivative as in (\ref{nome}).
We can now use Stokes theorem on the moduli space and reduce the integral to its boundary.
The boundary contribution is then given by the (finite) limit of the amplitude 
on the degenerate Riemann surface obtained by shrinking to the real codimension one
component of the moduli space. 
This was studied in the previous section where we described in detail its compactification
and its real codimension one boundary structure.
This is the relevant contribution for the open string moduli.

In order to calculate the boundary terms, we follow a technique similar to the one developed 
in \cite{bcov}, see Sects.3 and 4, although adapted to the present case.
A Riemann surface sitting in the neighborhood of the open boundary of the moduli space
$\partial_o {\cal M}_{g,h}$, see (\ref{open}), has a long strip which becomes a boundary
node in the degeneration limit $\varepsilon\to 0$ as in Fig.\ref{eight}.
We can choose coordinates near $\partial_o {\cal M}_{g,h}$ as $(\varepsilon, m', t_1, t_2)$
where $\varepsilon$ is the real plumbing fixture coordinate and $(m',t_1,t_2)$
are the moduli of the punctured Riemann surface resulting from the degeneration 
$\varepsilon\to 0$. In particular $(t_1,t_2)$ are the locations of the boundary punctures.
In the limit $\varepsilon\to 0$ the Beltrami differentials associated to the boundary
collision are supported near $(t_1,t_2)$ and their contribution to the measure
reads
\be
\int_{\gamma_{t_1}} (G^- + \bar G^-)\int_{\gamma_{t_2}} (G^- + \bar G^-) \ \ .
\ee
The contribution from $\partial_o {\cal M}_{g,h}$ to (\ref{nome}) is then 
\be
\int_{\partial_o {\cal M}_{g,h}} 
\langle \int_{\partial\sgh}\Theta_{\bar\alpha}
\int_{\gamma_{t_1}} (G^- + \bar G^-)\int_{\gamma_{t_2}} (G^- + \bar G^-)
\prod (m',G^-)
\rangle_{\sgh}
\label{A}
\ee
where $\prod (m',G^-)$ is the left-over measure factor corresponding 
to the moduli $m'$. 
\vspace{.5cm}
\begin{figure}[!h]
\begin{center}
\includegraphics{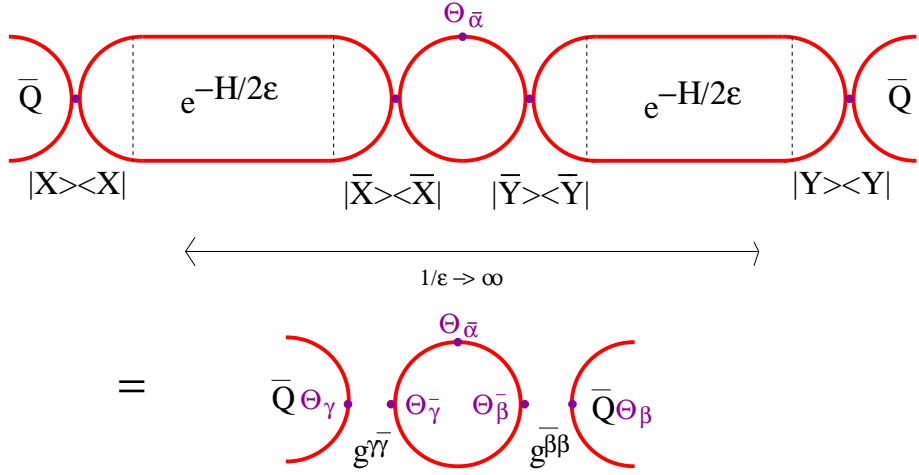}
\caption{the infinite strip contribution.}
\label{thestrip}
\end{center}
\end{figure}
Let we now rewrite the path integral on the long strip as depicted in Fig.\ref{thestrip}.
Namely, we
insert two chiral resolutions of the open string states identity $\sum_X|X\rangle\langle X|$ 
at the two ends and two anti-chiral ones  $\sum_{\bar X}|\bar X\rangle\langle\bar X|$
in the middle.
In the $\varepsilon\to 0$ limit, corresponding to the infinite 
length of the strip (see Fig.\ref{thestrip}), only the ground states do contribute and the
contributions of the two halves of the long strip give the open string metric insertions by definition.
Moreover, the only contribution to (\ref{A}) is when $\Theta_{\bar\alpha}$ is integrated along a boundary component 
involved in the degeneration limit, more precisely in the unit disk in the middle of the strip
(see once more Fig.\ref{thestrip}). In fact, when $\Theta_{\bar\alpha}$ is on a different boundary component, 
the amplitude is vanishing due to ghost number conservation.
Then (\ref{A}) becomes 
\be
\int_{\partial_o {\cal M}_{g,h}} 
\langle \Theta_\beta
\oint_{\partial{\Sigma_{0,1}}}
\Theta_{\bar\alpha}\Theta_\gamma\rangle_{{\Sigma}_{0,1}}
\langle \bar Q \Theta_\beta \bar Q \Theta_\gamma 
\prod (m',G^-)
\rangle_{{\Sigma}_{singular}}
\label{Y}
\ee
Actually, due to $PSL(2,\real)$ symmetry, we can fix all the three angular positions of the disk insertions.
The three-point function
\be
\Delta_{\bar\beta\bar\alpha\bar\gamma}=
\langle \Theta_\beta(-1)
\oint_{\partial{\Sigma_{0,1}}}
\Theta_{\bar\alpha}\Theta_\gamma(1)\rangle_{{\Sigma}_{0,1}}
\label{Delta}\ee
gives two contributions, 
corresponding to the two different orderings 
of three points on the disk boundary, which anti-symmetrize the two 
possible intermediate insertions.

The second factor in (\ref{Y}) can be rewritten
as two covariant derivatives of the topological string amplitude for the boundary Riemann
surface $\Sigma_{singular}$.

As it has been already discussed in Sect.3, the real codimension one component of the moduli space 
contributing to (\ref{uno}) includes also a component from Riemann surfaces obtained by shrinking 
to zero the length of the boundaries (see Fig\ref{shrinking}).
Therefore, on top of $\partial_o{\cal M}_{g,h}$, we have to consider the term 
${\cal M}_{g,h-1,n+1,\hat{\tt m}}\times {\cal M}_{0,1,1,m}$ in (\ref{cb}).
Near this boundary component, the Riemann surface develops a long tube.
\vspace{.5cm}
\begin{figure}[!h]
\begin{center}
\includegraphics{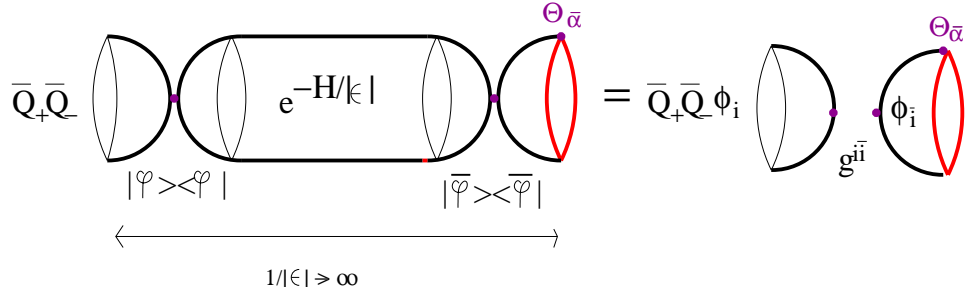}
\caption{the infinite tube contribution.}
\label{thetube}
\end{center}
\end{figure}
Let we now rewrite the path integral on the long tube as depicted in Fig.\ref{thetube}.
Namely, we
insert a chiral resolution of the closed string states identity $\sum_\varphi|\varphi\rangle\langle \varphi|$ 
at the beginning of the tube and an anti-chiral one  $\sum_{\bar \varphi}|\bar\varphi\rangle\langle\bar\varphi|$
at the end (see Fig.\ref{thetube}).
In the shrinking limit the amplitude gets projected on the chiral bulk ground states
and we therefore get
\be
\int_{{\cal M}_{g,h-1}}g^{\bar i i}\langle\oint_{\partial{\Sigma_{0,1}}}\Theta_{\bar\alpha} \phi_{\bar i}
\rangle_{{\Sigma_{0,1}}}
\langle \int_{{\Sigma_{g,h-1}}}\bar Q^+ \bar Q^- \phi_i \prod (m',G^-)\rangle_{{\Sigma_{g,h-1}}}
\label{421}
\ee
where the $\bar Q^+ \bar Q^-$ action is induced by the Beltrami differentials supported near the puncture.
As before, due to the $PSL(2,\real)$ invariance, the bulk to boundary disk function
$\oint_{\partial{\Sigma_{0,1}}}\langle\Theta_{\bar\alpha}\phi_{\bar i}\rangle_{{\Sigma_{0,1}}}$ 
is effectively  unintegrated.
The only contribution comes when the $\Theta_{\bar\alpha}$ insertion is along the boundary component at the end 
of the tube. The second factor of (\ref{421}) can be rewritten as the holomorphic derivative of the amplitude on 
$\Sigma_{g,h-1}$.

Summing the result of this manipulations we get the following four contributions corresponding to the 
pinching and dividing open paths connecting the same boundary component, the colliding
of two boundaries and the shrinking of the holes length respectively
$$
\partial_{\bar t_{\bar\alpha}}F_{g,h}=
\frac{1}{2}g^{\bar\beta\beta}g^{\bar\gamma\gamma}\Delta_{\bar\beta\bar\alpha\bar\gamma}
\left[
D_\beta D_\gamma F_{g-1,h+1} 
+ \sum_{{\tiny \matrix{g_1+g_2 &=& g\cr h_1+h_2 &=& h+1}}}
D_\beta F_{g_1,h_1} D_\gamma F_{g_2,h_2} 
+ D_\beta D_\gamma F_{g,h-1}
\right]
+
$$ 
\be
+g^{i\bar i}\Pi_{\bar\alpha\bar i} D_i F_{g,h-1}
\label{eccola}\ee
where
$g_{\alpha\bar\alpha}$ is the open string moduli metric (as in section 4 in \cite{bcov}) and 
\be
\Pi_{\bar\alpha\bar i}=\langle\Theta_{\bar\alpha}\phi_{\bar i}\rangle_{{\Sigma_{0,1}}}
\ee
is the overlap function.

Finally, in (\ref{eccola}),
$D_\alpha=\partial_\alpha-(2-2g-h)\partial_\alpha K_{open}- \Gamma_\alpha$ 
is the covariant derivative in the open string holomorphic moduli,
and $D_i=\partial_i-(2-2g-h)\partial_i K_{closed}- \Gamma_i$ is the covariant derivative 
in the closed string moduli.
The first term in the connection appears since $F_{g,h}$ is a section of the
${\cal L}^{2-2g-h}$ line bundle associated to the rescaling of the holomorphic three form
of the Calabi-Yau.
The two K\"ahler potentials are related to the vacua normalizations in the 
open and closed sectors. 

Let us remark that the open string amplitude $F_{g,h}$ is parametrized by the full
boundary chiral ring $H^{0,1}_{\bar\partial_A}\left(X,E^a\times {E^b}^*\right)$, where 
$E^a$ labels the different Chan-Paton indices associated to the branes \cite{hiv}.
Therefore the boundary insertions in the above disk amplitude
involve different chiral sectors corresponding to the specific boundary conditions for the open strings.
In particular, as it has been already observed in the explicit computations at genus $0$ in 
\cite{Marchesano,ant}, a non trivial holomorphic anomaly in the open string sector
can be present only if at least three different kinds of branes are involved.
Actually, this is necessary for the first disk contribution in (\ref{eccola})
not to vanish.
Notice that this result is in agreement also with calculations performed in local Calabi-Yau's
(\cite{open,vafa-o,marino,large}), where a single brane type appears and no holomorphic anomaly
in the open sector is observed.

\section{Closed moduli in presence of Wilson lines}

Let us now consider the variation of the closed string moduli in presence of non zero 
Wilson lines. This, as we explained in (\ref{mdef}), on top of generating bulk insertions
will add some boundary insertions mixing again open and closed moduli.
We calculate henceforth the variation of the topological string amplitude $F_{g,h}$ 
under an anti-holomorphic shift $w$ as in (\ref{mdef}). This gives
\be
\partial_{\bar t_{\bar i}}F_{g,h}=
\int_{{\cal M}_{g,h}} 
\langle\left(
\qbp\qbm\int_{\sgh} \phi_{\bar i}+
Q\oint_{\partial\sgh}\Psi_{\bar i}
\right)
\prod_{k=1}^{3g-3+h}|(\mu_k,G^-)|^2 \prod_{a=1}^h (\lambda_a,G^-)\rangle_{\Sigma_{g,h}}
\label{abc}\ee
where we used the notation introduced in (\ref{notation}).
For the sake of clarity, we split the calculation in the two additive factors in (\ref{abc}).

{\bf The first contribution in (\ref{abc})} has been already studied in \cite{jw} in the case in which the
Wilson lines were frozen. The same analysis can be repeated here with some care concerning 
the symmetries of the action (\ref{Bact}). In fact, in presence of the boundaries only the 
combinations $Q$ and $\bar Q$ are preserved implying that some new terms 
could arise once pulling the non conserved supercharges. 
Smeargingly, once we define the {\it nonconserved charge} $Q'=\oint (G^+-\bar G^+)$, 
the first term is
\be
\int_{{\cal M}_{g,h}} 
\langle\left(-\frac{1}{2}\right)
QQ'\int_{\sgh} \phi_{\bar i}
\prod_{k=1}^{3g-3+h}|(\mu_k,G^-)|^2 \prod_{a=1}^h (\lambda_a,G^-)\rangle_{\Sigma_{g,h}}
\label{closed1}\ee
While the charge $Q$ can be 
harmlessly pulled against the measure factor, the charge $Q'$ generates a new contribution proportional to 
$Q' S_B$. Notice that this is the integrated boundary insertion of the broken supercurrent $J'=G^+-\bar G^+$.
We thus get
\bea
&&\int_{{\cal M}_{g,h}} 
\langle\int_{\sgh} \phi_{\bar i} \left(-\frac{1}{2}QQ'
\prod_{k=1}^{3g-3+h}|(\mu_k,G^-)|^2\right) \prod_{a=1}^h (\lambda_a,G^-)\nonumber \\
&& + \int_{\sgh}\phi^{[1]}_{\bar i}\prod_{k=1}^{3g-3+h}|(\mu_k,G^-)|^2
\left(Q \prod_{a=1}^h (\lambda_a,G^-)\right) \rangle_{\Sigma_{g,h}} \label{mostruosa}\\
&&+ 
\int_{{\cal M}_{g,h}} \langle\int_{\sgh} \phi_{\bar i}\left(\frac{1}{2}\int_{\partial\sgh} J'\right)
\left(Q\prod_{k=1}^{3g-3+h}|(\mu_k,G^-)|^2\right)  \prod_{a=1}^h (\lambda_a,G^-) \rangle_{\Sigma_{g,h}}
\nonumber
\eea
where we defined $\phi^{[1]}_{\bar i}= \frac{1}{2} Q' \phi_{\bar {i}}$ and
we used the fact that the action of the non-conserved charge $Q'$ on the factor
of the measure containing the $\lambda_a$ differential is zero since it does not couple
to the real moduli.

The first two terms in (\ref{mostruosa}) give rise to the extended HAE studied in \cite{jw},
while the last term is a new contribution which we now calculate.
The degeneration of the Riemann surface associated with the action of $Q$ in the last term
of (\ref{mostruosa}) gives rise to a long strip and again this projects, 
as described in the previous section, on chiral boundary operators.
Because of ghost number conservation, the only contribution can come when both
$ \phi_{\bar {i}}$ and $J'$ are on the strip. We are then left with
\be
\frac{1}{2}g^{\bar\beta\beta}g^{\bar\gamma\gamma} B_{\bar\beta\bar i\bar\gamma}
\left[
D_\beta D_\gamma F_{g-1,h+1} + D_\beta D_\gamma F_{g,h-1} +
\sum_{{\tiny \matrix{g_1+g_2 &=& g\cr h_1+h_2 &=& h+1}}}
D_\beta F_{g_1,h_1} D_\gamma F_{g_2,h_2} 
\right]
\label{casinoso}
\ee
where 
\be
B_{\bar\beta\bar i\bar\gamma} = \int_0^{2\pi}\ d\vartheta\int_0^1\ dr 
\langle \Theta_{\bar\beta} (-1) J' (e^{i\vartheta}) \phi_{\bar i}(r)
\Theta_{\bar\gamma}(1)\rangle_{\Sigma_{0,1}}\ .
\label{semprepiucasinoso}
\ee 

{\bf The second term in (\ref{abc})} has exactly the same structure of (\ref{uno}) and therefore can be calculated
in full analogy with what we did in the previous section. This gives the following contribution
\be
\frac{1}{2}g^{\bar\beta\beta}g^{\bar\gamma\gamma}\Delta'_{\bar\beta\bar i\bar\gamma}
\left[
D_\beta D_\gamma F_{g-1,h+1} + D_\beta D_\gamma F_{g,h-1} +
\sum_{{\tiny \matrix{g_1+g_2 &=& g\cr h_1+h_2 &=& h+1}}}
D_\beta F_{g_1,h_1} D_\gamma F_{g_2,h_2} 
\right]
+
g^{j\bar j}\Delta'_{\bar i\bar j} D_jF_{g,h-1}
\label{seconda}\ee
where
\be
\Delta'_{\bar\beta\bar i\bar\gamma}
=
\langle\Theta_{\bar\beta}\oint_{\Sigma_{0,1}}\Psi_{\bar i}\Theta_{\bar\gamma}\rangle_{\Sigma_{0,1}}
\ee
and 
$\Delta'_{\bar i\bar j}=\langle\phi_{\bar j}\Psi_{\bar i}\rangle_{\Sigma_{0,1}}$.

Adding the two contributions of (\ref{abc}) we get the complete extended HAE for closed moduli which reads
\bea
&&\partial_{\bar t_{\bar i}}F_{g,h}=
\frac{1}{2} C_{\bar i}^{jk}\left[ \sum_{{\tiny \matrix{g_1 + g_2 &=& g\cr h_1 + h_2 &=& h}}}
 D_j F_{g_1,h_1} D_k F_{g_2,h_2} + D_j D_k F_{g-1,h} \right] 
- (\Delta + \Delta')^j_{\bar i} D_j F_{g,h-1} + \nonumber\\
&&+ \frac{1}{2} (\Delta' + B)^{\beta\gamma}_{\bar i}
\left[
D_\beta D_\gamma F_{g-1,h+1} + D_\beta D_\gamma F_{g,h-1} +
\sum_{{\tiny \matrix{g_1+g_2 &=& g\cr h_1+h_2 &=& h+1}}}
D_\beta F_{g_1,h_1} D_\gamma F_{g_2,h_2} 
\right]
\label{prova}
\eea
The indexes $(i,\alpha)$ of the closed and open moduli are raised as usual
via the (inverse) hermitian closed and open string metrics respectively.
 
Notice that switching off the Wilson lines at $A=0$ and declaring all the open moduli derivatives 
$D_\alpha$ to be zero at $A=0$ we recover as a sub-case the result in \cite{jw}.

\section{Open issues}

The main open issue is to understand the relationship between
the HAE's for open and closed moduli in the spirit of 
gauge/string duality. 
The similarity among the 
combinatorial structures of the
boundary of the compactified moduli spaces of Riemann surfaces
under the shrinking of open and closed paths as described in Section 3 
should play a full role in the solution of this open issue and could 
enlarge our knowledge about open/closed string duality.
In this context it will be crucial to develop a complete 
$tt^*$-geometry for open and closed moduli. This was analyzed in \cite{ttstar}
for the closed string and in \cite{bcov,herbst} for the open string. 
Actually, to our knowledge, the full $tt^*$-geometric structure is still uncovered.
Its geometrical data will include all the mixed (open and closed) correlators
entering our complete HAE's, and would provide a geometrical interpretation
to them. Our analysis is valid for Riemann surfaces with negative Euler
characteristic. As in the usual BCOV case, the other cases have to be studied
by direct inspection. In particular the holomorphic anomaly for the
annulus amplitude should be related to the Quillen anomaly 
\cite{bcov,large,jw},
while the bulk-to-boundary disk two point functions should 
have some relation to the Abel-Jacobi map \cite{vafa,jw}.
These specific correlators provide, up to the open moduli 
holomorphic ambiguity, the data needed to study the complete HAE's.
As an example one could study the particular case of the quintic
and the explicit form of our equations by implementing the complete
HAE's in the context of \cite{OOY} and \cite{BDLR}
\footnote{The dependence on the closed moduli has been analyzed at frozen open moduli in \cite{jw}.}.

The comparison of the B-model case, which we discussed in detail in this letter, and the A-model
could have some applications to mirror symmetry. This should be done with the due care
following the lines of \cite{vafa} and offers a relation to the open Gromov-Witten invariants
\cite{GZ}.

It is well possible that under some favorable conditions
one could find a suitable set of coordinates which simplifies
the structure of the HAE's by reabsorbing the dependence
on the open moduli by a shift of the closed ones or viceversa. 
A better comprehension of the general structure could be gained
by studying HAE's in a resummed form for the generating function
\break
$
{\cal F}(g_s,\lambda)=\sum_{g,h}F_{g,h} g_s^{2g-2+h}\lambda^h
$.
Actually, using this perspective it was shown in \cite{zio} that
the boundary effects in the closed HAE's
at frozen Wilson lines studied in \cite{jw} can be re-casted 
in a shift of variables of the closed string moduli. 
Moreover,
in the open string case we find that the HAE's do involve in
the right-hand-side terms with
higher number of boundary components, although with lower genera. 
As such they do not
admit an interpretation as recursive relations in the genus 
equal to the one found in the closed string case \cite{bcov}.
The correct quantity to be considered in the case of open Riemann surfaces 
is instead $\chi_{\sgh}=2-2g-h$ which increases
passing to the moduli space boundary components (while the stabilized Euler characteristic
(\ref{euler}) of Section 3 is of course invariant).
Namely, the HAE's relate the anti-holomorphic derivatives of the $F_{g,h}$
to the holomorphic derivatives of the same objects with lower $2g+h-2$ 
(and, needless to say, not increasing genus).
It would be very useful to explore this point in further detail
in order to understand the resolvability of the complete HAE's.

The analogue of the analysis in \cite{marino} should hold for our HAE's too, 
by mapping them to loop equations for suitable matrix models.
Notice that since we included in our analysis non-trivial boundary states,
we expect our equations to be viable also for the analysis of local Calabi-Yau's,
by properly taking into account the presence of a non-trivial superpotential
which modifies the boundary chiral ring.
 
Restricting our results to genus zero one should reproduce \cite{Marchesano,ant,Rodolfo}. 
Actually one can check that our equations reduce for $g=0$ to the ones obtained in 
\cite{ant} after a suitable interpretation of peculiar operatorial insertions
\footnote{
Specifically, 
our equations (\ref{eccola}) and (\ref{prova}) reduce for genus zero 
to eqs.(4.6-8) in \cite{ant}, 
the major difference being that we obtained an explicit expression for 
the amplitudes with anti-chiral insertions in \cite{ant}
in terms of derivatives of chiral amplitudes contracted with
open/closed string metric and disk functions. This allowed us to write a closed
system of equations.}.
In \cite{ant} it was observed that the amplitudes $F_{g,h}$ for $g>0$ 
do not have a straightforward interpretation as F-terms in a four-dimensional 
Poincar\'e invariant superstring compactification on $R^4\times$CY.
The low-energy limit of these amplitudes can be nonetheless interpreted
as the superpotential of the four dimensional ${\cal N}=1$ field theory
living on space-time filling branes wrapped on internal cycles of the 
(non-compact) CY \cite{vafa-j}. 

The analysis of our equations could clarify some
issues on the 
(non-)holomorphicity of these superpotentials
which arise in the study of intersecting brane models \cite{Marchesano}. 
To this end, one should also generalize the analysis presented in this letter
to the non-abelian case and study in detail the boundary
conditions for the open strings in presence of different stacks of branes. 
The outcome should be a tensorization of our HAE's with 
the Lie algebra of the Chan-Paton factors and the boundary condition mixing.

We hope to come back to some of the above open issues in future publications.

{\bf Acknowledgments}:
We thank B.Fantechi, M.Marino, H.Ooguri, R.Russo and J.Walcher for discussions and exchange of opinions.
The research of G.B. is supported by the European Commission RTN Program MRTN-CT-2004-005104 and by MIUR.

\end{document}